# Superconducting properties of FeSe wires and tapes prepared by gas diffusion technique

Zhaoshun Gao, Yanpeng Qi, Lei Wang, Dongliang Wang, Xianping Zhang, Chao Yao, Yanwei Ma*

Key Laboratory of Applied Superconductivity, Institute of Electrical Engineering, Chinese Academy of Sciences, Beijing 100190, China

**Abstract:**

Superconducting FeSe in the form of wires and tapes were successfully fabricated using a novel gas diffusion procedure. Structural analysis by mean of x-ray diffraction shows that the main phase of tetragonal PbO-type FeSe was obtained by this synthesis method. The zero resistivity transition temperature of the FeSe was confirmed to be 9.3 K. The critical current density as high as 137 A/cm$^2$ (4 K, self field) has been observed. The results suggest that the diffusion procedure is promising in preparing high-quality FeSe wires and tapes.

* Author to whom correspondence should be addressed; E-mail: ywma@mail.iee.ac.cn

The discovery of superconductivity in the iron pnictides has triggered great interest in past three years [1-7]. In addition to the high transition temperature, $T_c$, these Fe based superconductors were reported to have a very high upper critical field, $H_{c2}$, bringing the hope in a wide array of future applications [8–10]. Up to now, several groups of iron-based superconductors have been discovered, such as LnOFeAs (1111 series) [1-3], BaFe$_2$As$_2$ (122 series) [5], LiFeAs (111 series) [11-12] and FeSe (11 series) [6]. Among them, the 11 series have great advantage for application due to the simple structure and no toxic arsenic. Very recently, by intercalating alkali metals into between the FeSe layers, superconductivity around 30 K has been achieved [13-17]. Therefore, the FeSe based materials deserve intensive studies for both fundamental physics and potential applications.

The common procedure for producing FeSe wires is powder in tube (PIT) method [18, 19]. However, the PIT process is difficult to achieve high-density FeSe due to the package or the shrinkage of core materials. Hence, the transport $J_c$ of FeSe wire fabricated by this method is very low. An efficient technique to fabricate high-quality FeSe wires or tapes is still needed. As we know, diffusion technique has been widely used for fabrication of superconducting tapes or wires such as Nb$_3$Sn [20], MgB$_2$ [21, 22] and so on. In this paper, a novel gas diffusion procedure was developed for preparing high-density FeSe superconductor.

The samples were prepared in the following way. The Se powder and the pure iron wires and tapes were used as raw materials. After cleaning carefully, the iron wires and tapes were cut into 4~6 cm and sealed into a Fe tube with proper amounts of Se powder. Then these wires and tapes were reacted with Se vapor at temperatures varying between 400°C and 800°C. The reaction time varied between 8 h and 12 h. The final product are wires and tapes with a FeSe layer ranged from several microns to about 0.1 mm thick (Fig.1), depending on the treatment process and the amounts of Se powder. The FeSe layer can be easily separated from iron wires and tapes. For further investigation, all studied samples are taking from the tape's surface.

The phase identification and crystal structure investigation were carried out using x-ray diffraction (XRD). Temperature dependence of resistivity and transport critical current density $J_c$ were measured by standard four-probe method using a physical property measurement system (PPMS). Microstructure was studied using a scanning electron microscopy (SEM).

Fig.2 shows X-ray diffraction patterns for FeSe samples. Polycrystalline FeSe was obtained as a main phase with tetragonal PbO-type structure except for small amounts of impurity phases described as asterisks. The hexagonal-NiAs-type FeSe was not observed in our samples.

Fig.3 displays the temperature dependence of electrical resistivity for FeSe samples with the measuring current of 1mA. From this figure, a sharp drop in resistivity was observed below the onset temperature of about 15.1 K, and zero resistivity is attained below 9.3 K. This value is higher than the onset transition temperature for FeSe prepared by solid state reaction method [6]. The normal state resistivity of our sample shows a broad bump around 250 K and exhibits metallic behavior below 250 K. The similar behavior was also observed in FeSe samples [6]. The superconductivity of our sample is also confirmed by DC magnetization measurement which is shown in inset Fig.3. The relatively broad magnetic suggests that the inhomogeneous still present in our sample.

In order to investigate the microstructure of our samples, the SEM microanalysis is employed in fig.4. It can be seen that the FeSe layer has a very dense structure with few cracks. The cracks are probably original from the difference in thermal expansion between FeSe layer and iron. From the higher magnification images we can clearly see a layered structure, very similar to what has been observed in other iron based superconductors [10]. It is worth to note that the textural structure was detected in some samples as shown in Fig .4c.

Transport critical current density $J_c$ as a function of temperature is presented in Fig.5. A transport $J_c$ as high as ~137 A/cm$^2$ at 4 K and self field has been observed in FeSe superconductors, much higher than the value of Fe(Se,Te) wire fabricated by PIT method [18, 19], which have $J_c$ values only 12.4 cm$^2$ and 64.1 A/cm$^2$, respectively.

The higher $J_c$ values of the gas diffusion processed wires are presumably due to the higher FeSe layer density and the textural structure, as shown in Fig.4. However, the $J_c$ value of our sample is still lower than that obtained in the case of single crystal which generally attained $10^5$ A/cm$^2$ at 5 K [23]. Clearly, the impurity phases and the micro cracks, which were observed in the XRD pattern and SEM image, were thought to be harmful to the transport capability. We expect that the transport $J_c$ can be more increased by reducing the impurity phases, improving texture or introducing the flux pinning centers.

Fig.6a shows the resistive superconducting transitions for the FeSe sample under various magnetic fields. We tried to estimate the upper critical field ($H_{c2}$) and irreversibility field ($H_{irr}$), using the 90% and 10% points on the resistive transition curves. The temperature dependence of $H_{c2}$ and $H_{irr}$ with magnetic fields up to 9 T for the FeSe samples are determined in this way and shown in Fig. 6b. It is clear that the curve of $H_{c2}$ (T) is very steep with a slope of $- dH_{c2}/dT|_{Tc} = 3.33$ T/K, which significantly exceeds the Pauli limit 1.84 T/K. The excellent $H_{c2}$ (T) property indicates that this superconducting wire has an encouraging application in high fields.

We have prepared FeSe wires and tapes via the exposure of iron wires and tapes to Se vapor. The superconductivity with the zero resistivity transition temperature of 9.3 K was obtained. The transport $J_c$ as high as ~137 A/cm$^2$ at 4 K and self field has been observed The $J_c$ value has much potential to be improved by perfect texture or optimization of fabrication process. Our results clearly demonstrate that this synthesis technique is unique and simple, hence will be able to be applied to the fabrication of the other iron-based superconducting wires such as FeAs 122 series or recently discovered FeSe 122 series.

This work is partially supported by National '973' Program (Grant No. 2011CBA00105) and the National Natural Science Foundation of China (Grant No. 51002150 and 51025726).

# Captions

Figure 1  Photograph of the final FeSe wires and tapes.

Figure 2  XRD patterns of FeSe Layer after peeling off the iron substrate. The impurity phases are marked by *.

Figure 3  Temperature dependence of resistivity for FeSe sample at zero field up to 300 K. The inset shows the temperature dependence of dc magnetization for ZFC and FC processes at a magnetic field of H = 20 Oe.

Figure 4  SEM micrographs for the FeSe layers after peeling off the iron substrate. The cracks were indicated by arrows in the fig.4a.

Figure 5  Zero-field current-voltage characteristics at indicated temperatures for the FeSe sample.

Figure 6  (a) Temperature dependence of resistivity under magnetic fields up to 9 T for the FeSe sample. (b) The temperature dependence of $H_{c2}$ and $H_{irr}$ determined from 90% and 10% points on the resistive transition curves.

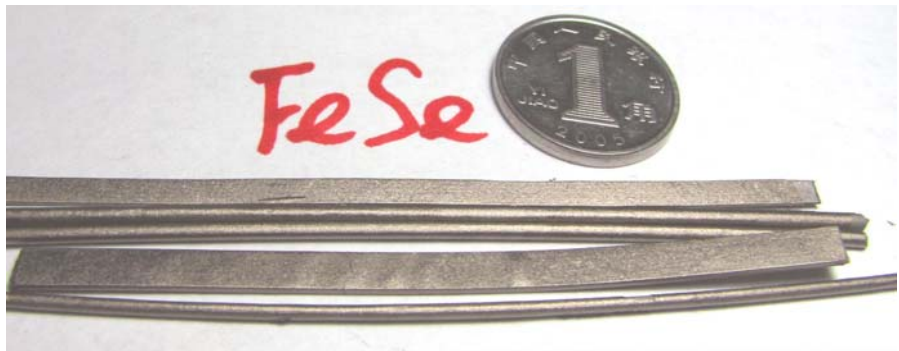

Fig.1 Gao et al.

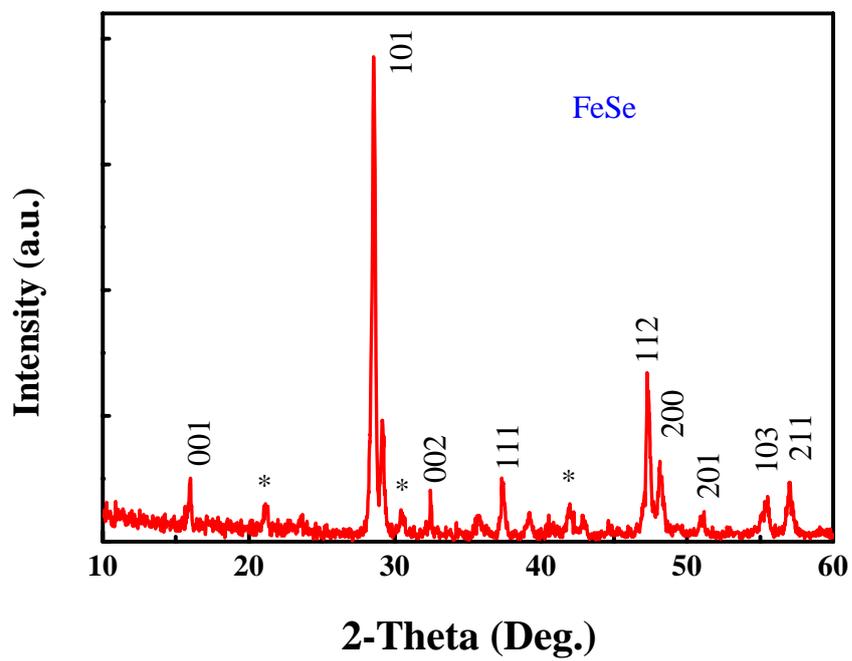

Fig.2 Gao et al.

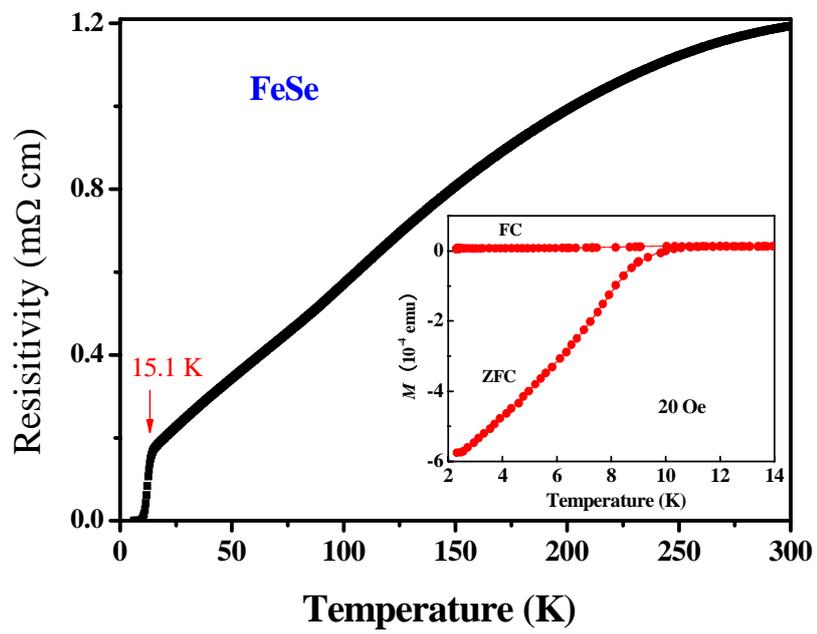

Fig.3 Gao et al.

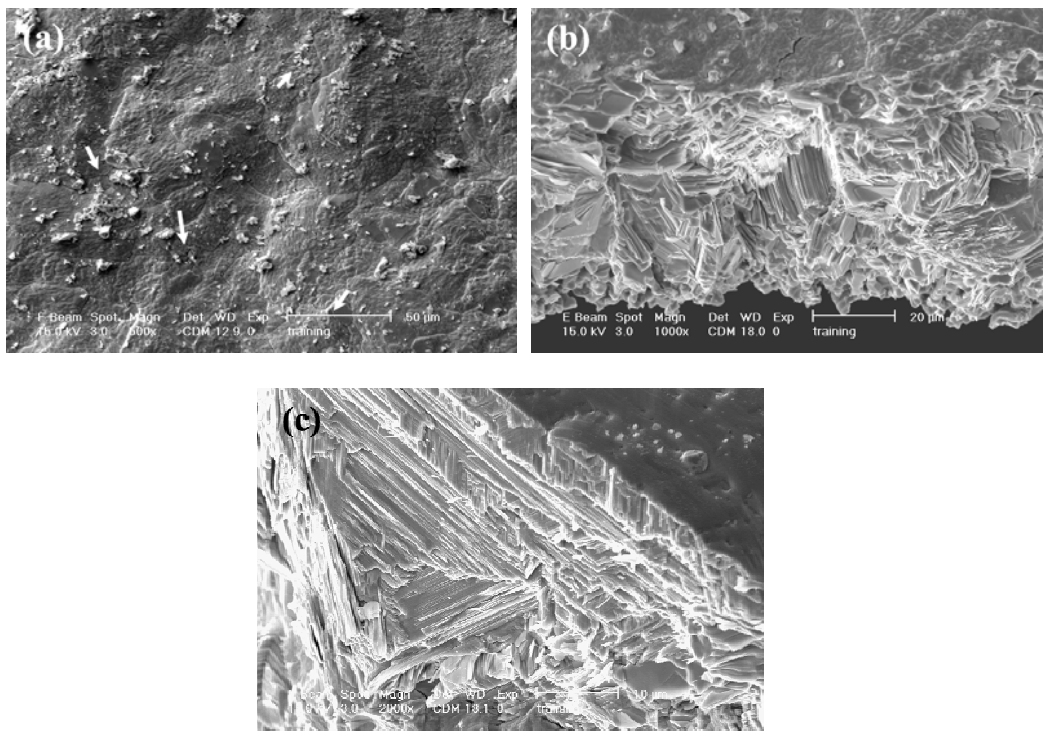

Fig.4 Gao et al.

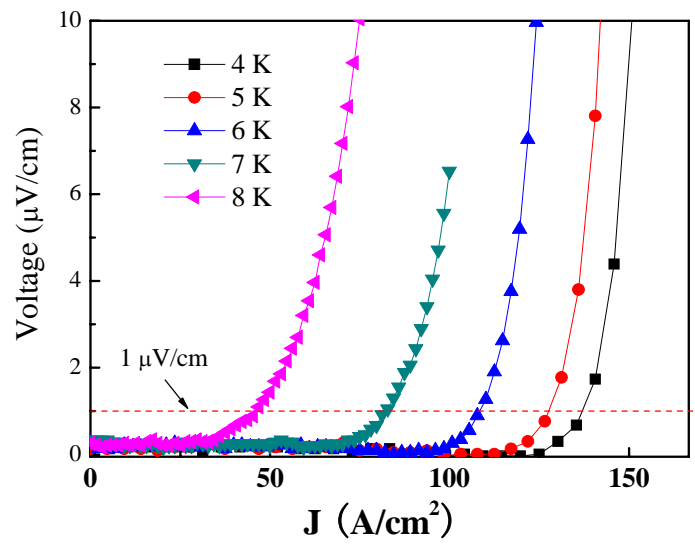

Fig.5 Gao et al.

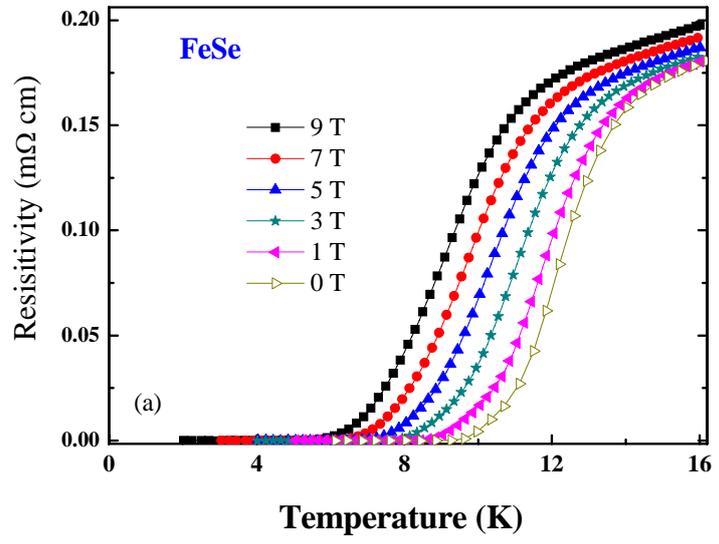

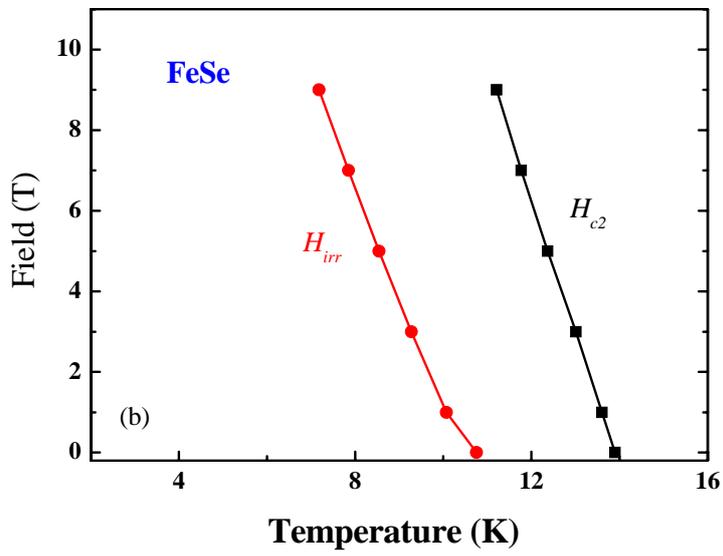

Fig.6 Gao et al.